The sustainome of global goal interactions varies by country income and is disproportionately influenced by inequalities.

David Lusseau[1] & Francesca Mancini[1]

[1]University of Aberdeen, School of Biological Sciences, Aberdeen AB24 2TZ, UK

Corresponding author: David Lusseau, Tel: +44 1224272843, Email: d.lusseau@abdn.ac.uk

**Abstract**

We interact with each other and our environment in rich and complex ways. These interactions form socioecological systems in which human, economic, or natural resources are used and replenished. In 2015, the United Nations set seventeen sustainable development goals (SDGs) to attempt to change the way we live and create by 2030 a sustainable future balancing equitable prosperity within planetary boundaries. We have tended to tackle SDGs in isolation and now we realise that a key hurdle to SDG implementation are conflicts arising from SDG interactions. We estimate here for the first time the sustainome, a global picture of those interactions, and determine the main hurdles to maximising SDG implementation. We show that the relative contribution of SDGs to global sustainable success differ by country income. SDG conflicts within the sustainome mean that we must find new ways to address the impacts of climate change, approaches to reducing inequalities and responsible consumption. Focussing on poverty alleviation and reducing inequalities will also have compounded positive effects on the sustainome. This network approach to sustainability provides a way to prioritise SDG and contextualise targets.

Many conflicts are emerging from the way we interact with each other and with our planet[1]. Since 1992, we have embarked on a range of global initiatives to find a more sustainable and equitable solution to these interactions. In 2015, we set a 15-year plan [2], composed of 17 sustainable development goals and 169 associated targets, to promote prosperity for all while protecting our planet [3]. Those goals touch on all aspect of human life and therefore interact in complex ways. We know that goal interactions can lead to conflicts and barriers to our plan [2,4]. While we are investigating interactions between some goal pairs, we do not have a complete picture of those global interactions to date. This is important as efforts to meet SDGs in isolation may be counter-productive [5] if they affect other SDGs negatively. Not all interactions may be negative and therefore we may also play on SDG effects on others for our efforts to have compounded effects on multiple goals. Studying the topology and drivers of interaction networks has given us crucial insights in the study of complex systems such as health [6,7], ecosystems [8] and our societies [9]. Here we propose to take a similar network approach to the SDGs and estimate the sustainome, the system of SDG interactions. We propose to do so at two scales: estimating interactions among the 169 SDG targets, and estimating interactions among the 17 SDG themselves.

The World Bank has developed an initiative to inform the SDGs with 331 indicators they have been regularly collecting over the past 27 years for 263 countries [10]. These indicators have been assigned to SDG targets and we can therefore see these indicators as measures of SDG progress, and by consequence SDG target progress too. We used these indicators to inform interactions between their respective identified 71 targets and 17 SDGs. Pairwise interactions between targets and SDG were estimated using pairwise meta-analyses of the standardised coefficients of association between the relevant indicators (54615 indicator association mixed effects models, 2556 and 153 meta-analyses for targets and SDGs respectively – seem methods for details). We therefore obtained two undirected (as we estimated associations), signed (positive and negative associations) weighted networks, one with 71 nodes (targets) and another with 17 nodes (SDGs) (Fig. 1).

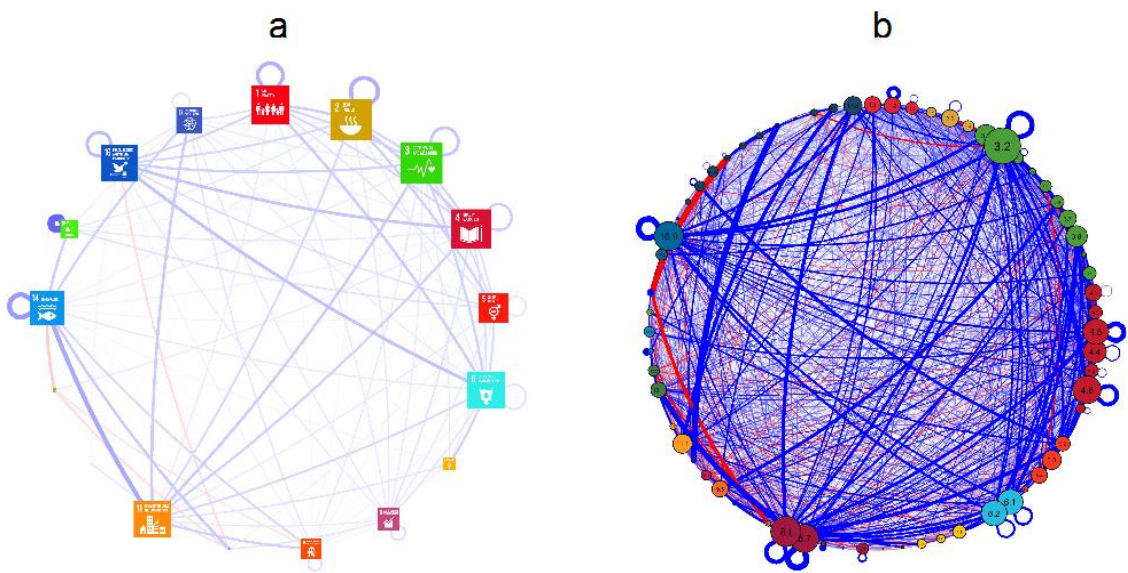

**Fig. 1.** SDG (a) and target (b) sustainome including all countries. Nodes are targets or SDGs and edges are associations (positive in blue and negative in red) with the line thickness representing the magnitude ranging from -1 to 1. Node size corresponds to the target or SDG eigenvector centrality highlighting the structural importance of each node.

We could use their respective graph Laplacian to understand the network overall dynamics [11]. Both sustainome networks are unstable and composed of antagonistic subgroups (target: 8 positive eigenvalues, SDG: 2 positive eigenvalues). In the case of SDGs, (SDG10,SDG12) and SDG13 are two clusters that are separate from the others. Those are also goals that have more negative links than positive in the network (Fig. 2) and are more likely to challenge our ability to meet all SDGs.

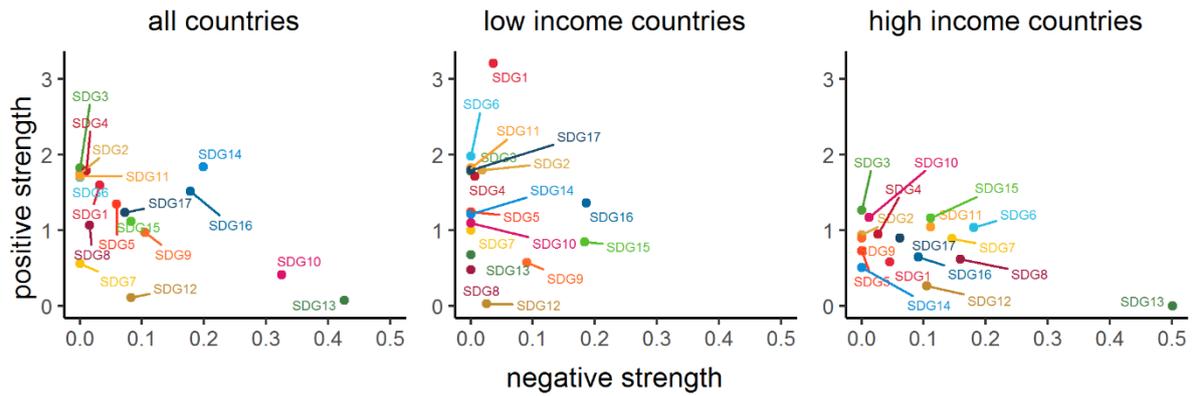

**Fig. 2.** Topological centrality of SDG depending on the sum of their positive (positive strength) and negative (negative strength) associations when considering all countries, low income countries and high income countries. Colors correspond to the SDG colours.

Level of income is a major macroeconomic driver of countries [12,13]. Therefore, we replicated this approach across the four income groups of countries defined by the World Bank [14] to determine whether the sustainome topology varied by income. We find that the topology of the sustainome changes drastically across income category (Figs. 3a & 4a; SI Appendix Figs. S1&S2). In those networks, some goals ('No Poverty' for low income, 'Reduce Inequalities' for high income) emerge as clear structural priorities for the SDG sustainome (Figs. 2, 3a & 4a). The behaviour of the sustainome also reacts to small changes in interactions for these goals (Fig. 3b & 4b).

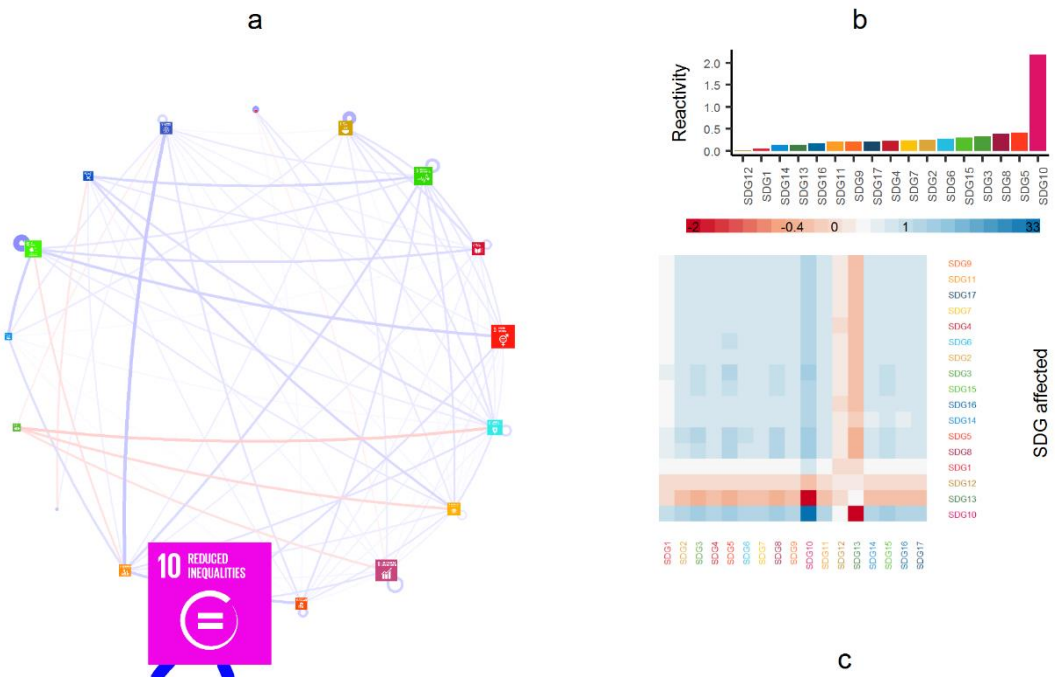

**Fig. 3.** The sustainome for high income countries (a), the contribution of each SDG to the reactivity of this sustainome (b) and the fate of all SDGs ($a_{1000}$, rows) as we intervene on each given SDG (column) given interactions in the sustainome (negative values correspond to a degradation of the SDG).

The SDG sustainome of low income countries does not contain antagonistic groups (all eigenvalues of the Laplacian ≤0, Fig. 4c). It appears therefore that our current goals will align in those countries and interventions in one goal is not likely to impede our ability to progress on other goals (Fig. 4c). To further understand this behaviour, we simply simulated the behaviour of SDGs (see methods, Figs. 3c & 4c, SI Appendix Figs. S1&S2) and targets (SI Appendix Fig. S3) as we intervened on others. For high income countries, SDG13 ('climate actions') is at odd with other goals (Figs. 2 & 3c) and interventions on SDG10 ('reduce inequalities') has the most positive effect on all other goals (Fig. 3c).

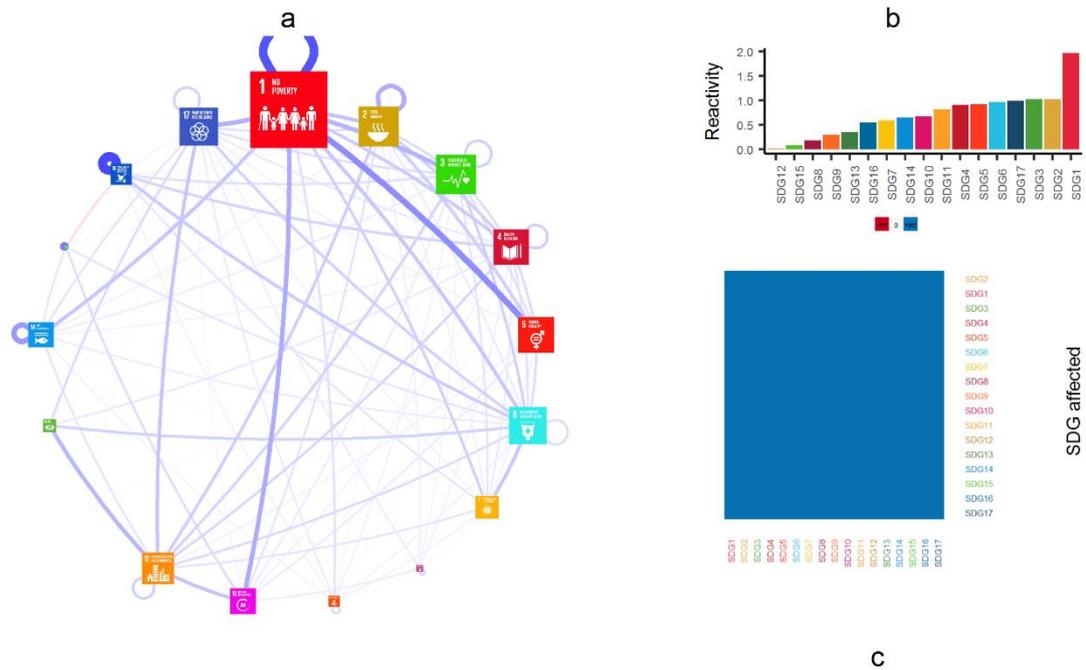

**Fig. 4.** The sustainome for low income countries (a), the contribution of each SDG to the reactivity of this sustainome (b) and the fate of all SDGs (rows) as we intervene on each given SDG (column) given interactions in the sustainome (negative values correspond to a degradation of the SDG).

The picture at the target level is more complex (SI Appendix Fig. S4). Target 3.2 ('reduce child mortality') remains a structurally-important component for SDG implementation (SI Appendix Fig. S4), its interaction pattern with other targets does not change with income (negative v positive strength, SI Appendix Fig. S5), and the target sustainome for all income groups will react more to changes in a reduction in child mortality (SI Appendix Fig. S6). When we focus on those interactions impeding SDG implementation (negative edges, SI Appendix Fig. S7) we see that this network subset differs widely across income, with health being replaced by a wider diversity of goals as being important barriers for high income countries; particularly climate challenges.

**Conclusions**

Our understanding of the sustainome will evolve as more data becomes available not only to enrich existing indicators, but also to define clear global indicators for the 98 targets currently unmonitored. Addiontal data will also enable us to use causal inferential frameworks as the current data sparsity constrained us to the current approach [15]. Analysing the sustainome provides us with new insights in the best way to achieve as many of our goals as possible by 2030. Firstly, prioritising reducing children early deaths will have compounded positive effects on the other targets. Secondly, we should contextualise our targets and prioritise our goals by country income levels. Prioritising poverty alleviation in low income countries and reducing inequalities in high income countries will have compounded positive effects on all SDGs. For high income countries, combatting climate change present the most barriers to achieving other goals. We therefore must change the way we are engaging in this climate challenge compared to the way we have approached it so far. This was echoed by the work presented for the Paris Agreement [16]. The sustainome of low income countries appears currently as the most achievable. The SDG sustainome did not scale up from the target interaction network, as what appear as structurally important targets were not necessarily included in structurally important goals in the sustainome. Staying flexible on targets but remaining focussed on goals appears to offer more opportunities to avoid SDG conflicts and achieve overall sustainability across contexts for each country.

**Materials and Methods**

**Estimating target and SDG networks.** We used World Bank indicator time series to inform SDG targets. The World Bank has developed a data bank of indicators categorised by the targets for which they are relevant [10]. We used 331 indicators, collected from 1990 to 2017, to inform 71 targets for all 17 SDGs. Original time series of natural disasters summarized for target 13.1 were retrieved from the EM-DAT [17]. We estimated the association between each indicator pairs using linear mixed effect models (MEMs) with country of observation origin as a random effect and an autoregressive correlation structure with a lag of one year within countries (using nlme in R [18]). Indicators were centered on their mean and scaled by their variance to obtained standardised coefficient of association assuming a gaussian distribution for the residuals; which was validated given observed residual distributions. Prior to fitting MEMs, scaled indicators were also informed by target directionality. The scaled times series were multiplied by -1 if the indicator definition was opposed to the desired trajectory and 1 if they were concomitant. For example, if our target is a reduction in child mortality and the indicator report the number of children that died annually in a country, we would like to see a decrease in this indicator. That way, when we estimated the association of indicator pairs we could use the standardised coefficients of associations ($\beta$s) to estimate whether indicator interactions are contrary or not to our targets. $\beta<0$ will indicate that the association is undesirable given our targets while $\beta>0$ will indicate that the association moves indicators in the desired direction. To estimate target interactions and SDG interactions we used these standardised coefficients ($\beta$), with their associated standard errors (SE) and meta-analysed these effects for each target and SDG pairs given the target and SDG membership of each indicators involved in the MEMs respectively. A pair of indicators was only considered once in these meta-analyses, which were mixed effect models of the standardised coefficients with a constant fixed effect (the interaction estimate) weighted by the SEs of the $\beta$s (using metafor in R [19]). Resulting estimated association coefficients ($\hat{\beta}$) significantly different from zero (p<0.05) were retained to estimate the signed weighted network **A** so that:

$\mathbf{A}_{ij} = \mathbf{A}_{ji} = \hat{\beta}$ which was a 71 x 71 matrix for targets and a 17 x 17 matrix for SDGs.

**Network state.** We used the graph Laplacian **L** of the signed networks **A** [11] to determine network stability (stable when all eigenvalues of the graph Laplacian ≤0): $L_{ij} = \begin{cases} L_{ij} = A_{ij} \; i \neq j \\ L_{ij} = -\sum_k A_{ik} \; i = j \end{cases}$. If the network was unstable we determined the number of antagonistic clusters in the networks using simulations. We simulated the vector of target/SDG states through time (1000 steps), **a**(t) so that:

$\mathbf{a}_{t+1} = \mathbf{A} \cdot \mathbf{a}_t$ and $\mathbf{a}_0 = [1]^n$ where n is the respective dimension of **A** (17 for SDG and 71 for targets). We repeated this simulation 17 times for SDG and 71 times for targets and for each set intervene on a single goal/target, i, at each time step ($\mathbf{a}_{i,t}=\mathbf{a}_{i,t}+0.1$) before estimating $\mathbf{a}_{t+1}$.

**Contributions of targets and SDGs to the sustainome.** We then estimated the partial contribution of each target and sdg in their respective signed networks to the reactivity [20] of the network to perturbation. This provides an understanding of the relative contribution of each target and sdg to the sustainome dynamics as we attempt to drive it towards our goals. Finally, we estimated the eigenvector centrality of each node in their respective networks as well as their positive and negative strengths to understand their relative contribution to the sustainome topology.

**Income group level estimations.** This network analytical process, including network estimation, was replicated for subset of countries categorised by their income using World Bank categories. We determined whether sustainome findings were consistent across income groups.

**Acknowledgements.** We thank the United Nations Department of Public Information for making the 17 SDG icons available. We thank the World Bank for curating, collating, and making the SDG indicator data easily accessible. We thank Alex Douglas for fruitful analytical discussions.

**Author Contributions.** DL conceived the study. DL and FM designed the statistical models, DL carried out the analyses and DL and FM wrote the manuscript.

**Author Information**. DL and FM have no competing interests associated with this work.

**Supplementary Information Appendix**

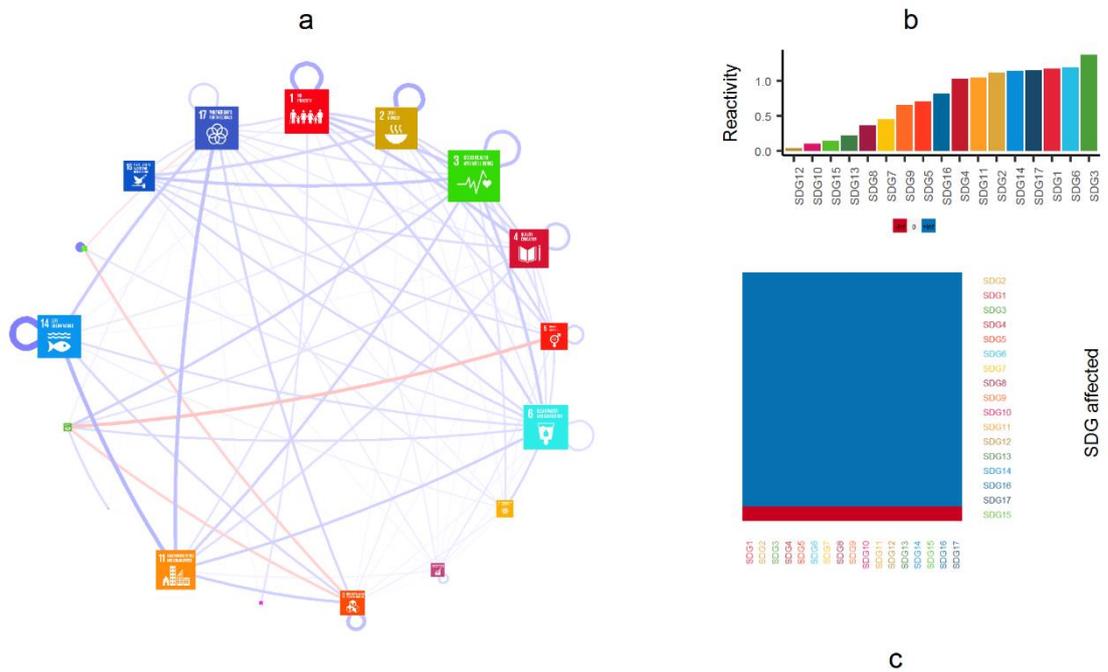

Fig. S1. The sustainome for low middle income countries (a), the contribution of each SDG to the reactivity of this sustainome (b) and the fate of all SDGs ($a_{1000}$, rows) as we intervene on each given SDG (column) given interactions in the sustainome (antagonistic groups of SDGs diverge towards +Inf and -Inf).

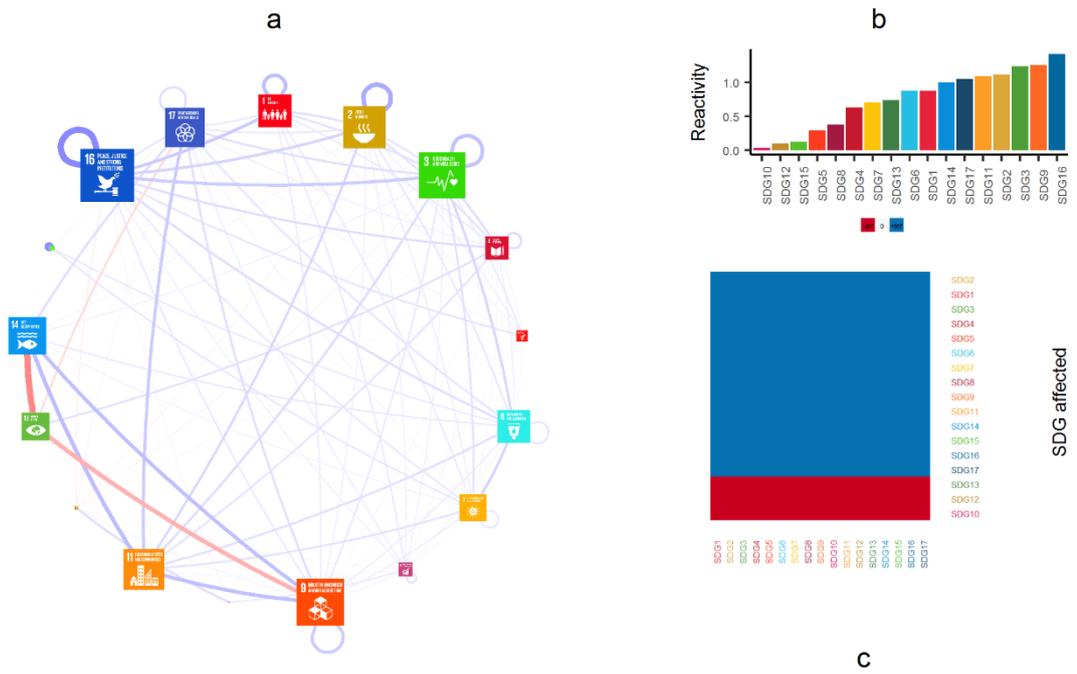

Fig. S2. The sustainome for high middle income countries (a), the contribution of each SDG to the reactivity of this sustainome (b) and the fate of all SDGs ($a_{1000}$, rows) as we intervene on each given SDG (column) given interactions in the sustainome (antagonistic groups of SDGs diverge towards +Inf and -Inf).

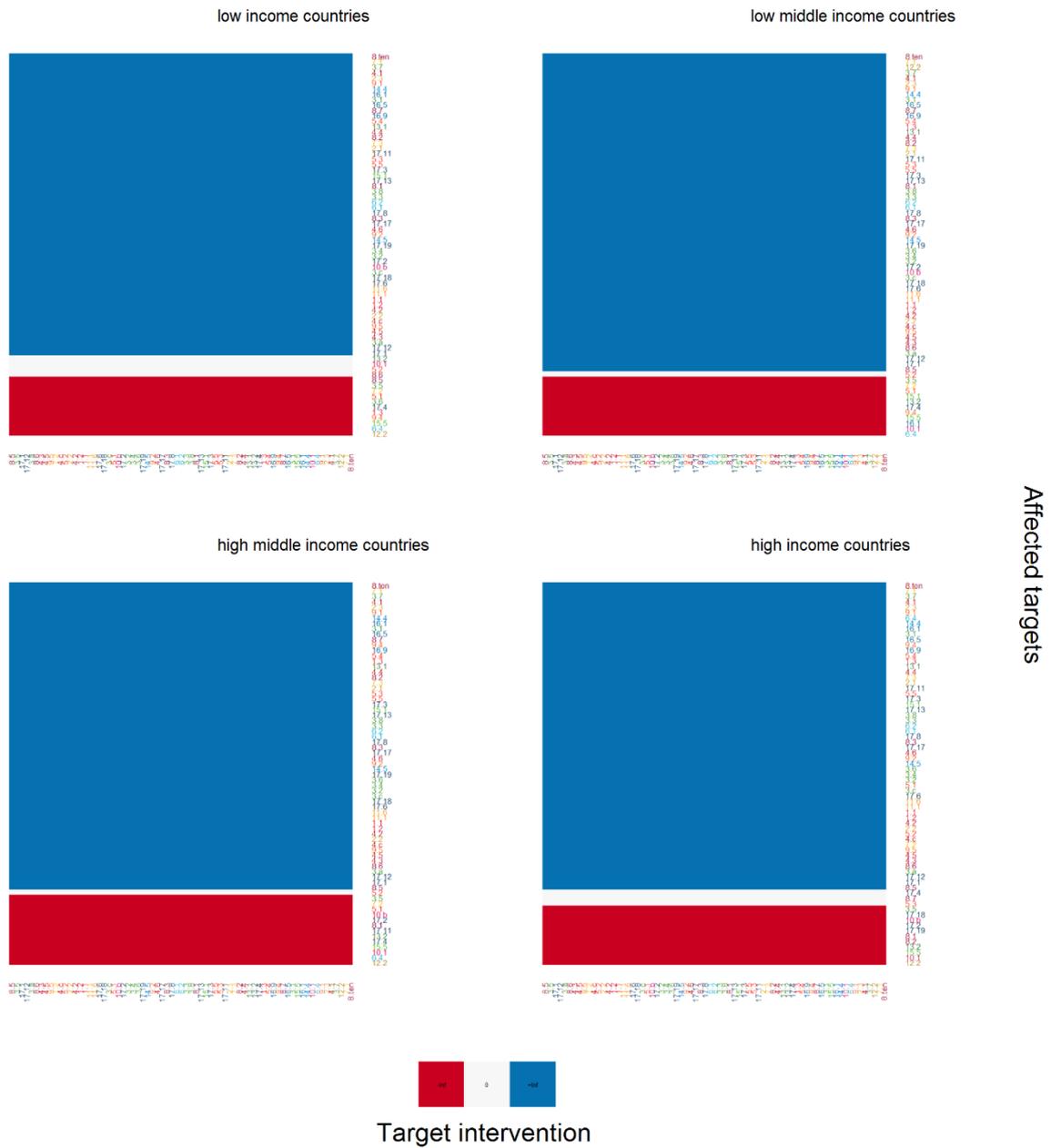

Fig. S3. the fate of all targets ($\mathbf{a}_{1000}$, rows) as we intervene on each given target (column) given interactions in the target sustainome. Antagonistic groups of targets are identified by their direction through time (+Inf, zero or -Inf).

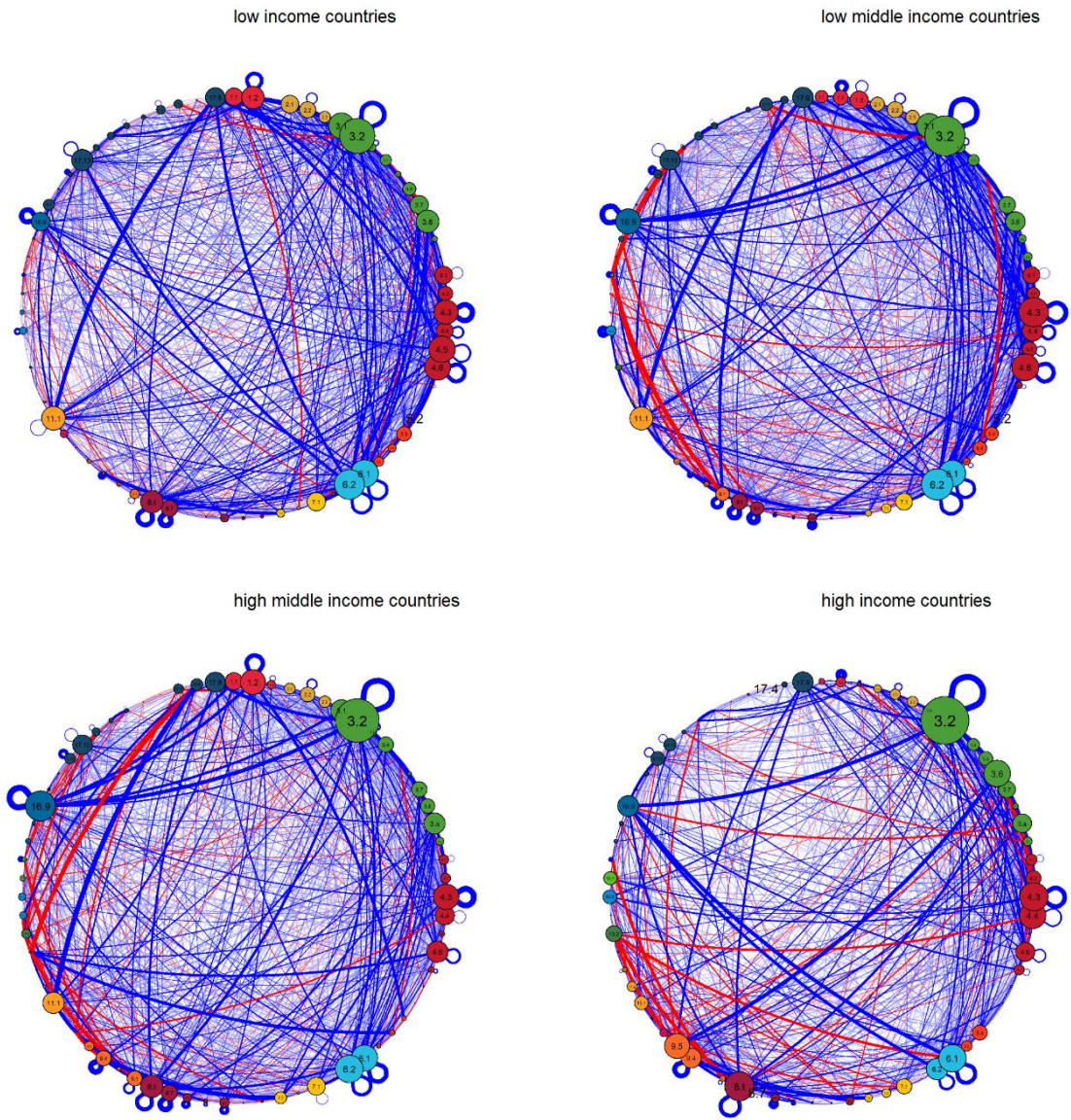

Fig. S4. Target sustainome for the four income categories. Nodes are targets or SDGs and edges are associations (positive in blue and negative in red) with the line thickness representing the magnitude ranging from -1 to 1. Node size corresponds to the target eigenvector centrality highlighting the structural importance of each node.

Fig. S5. Positive and negative strength for each target in the four target sustainomes.

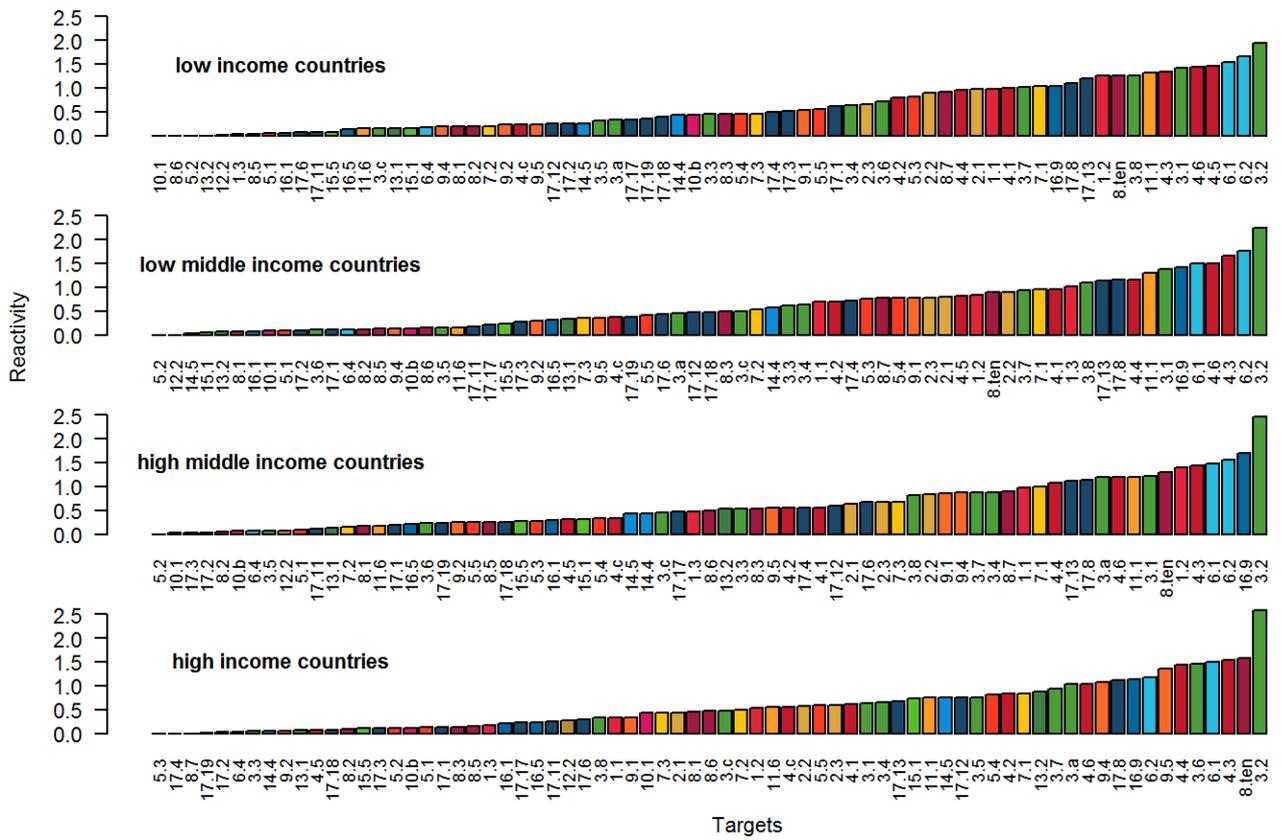

Fig. S6. Contribution of each target to the reactivity of each of the four target sustainomes.

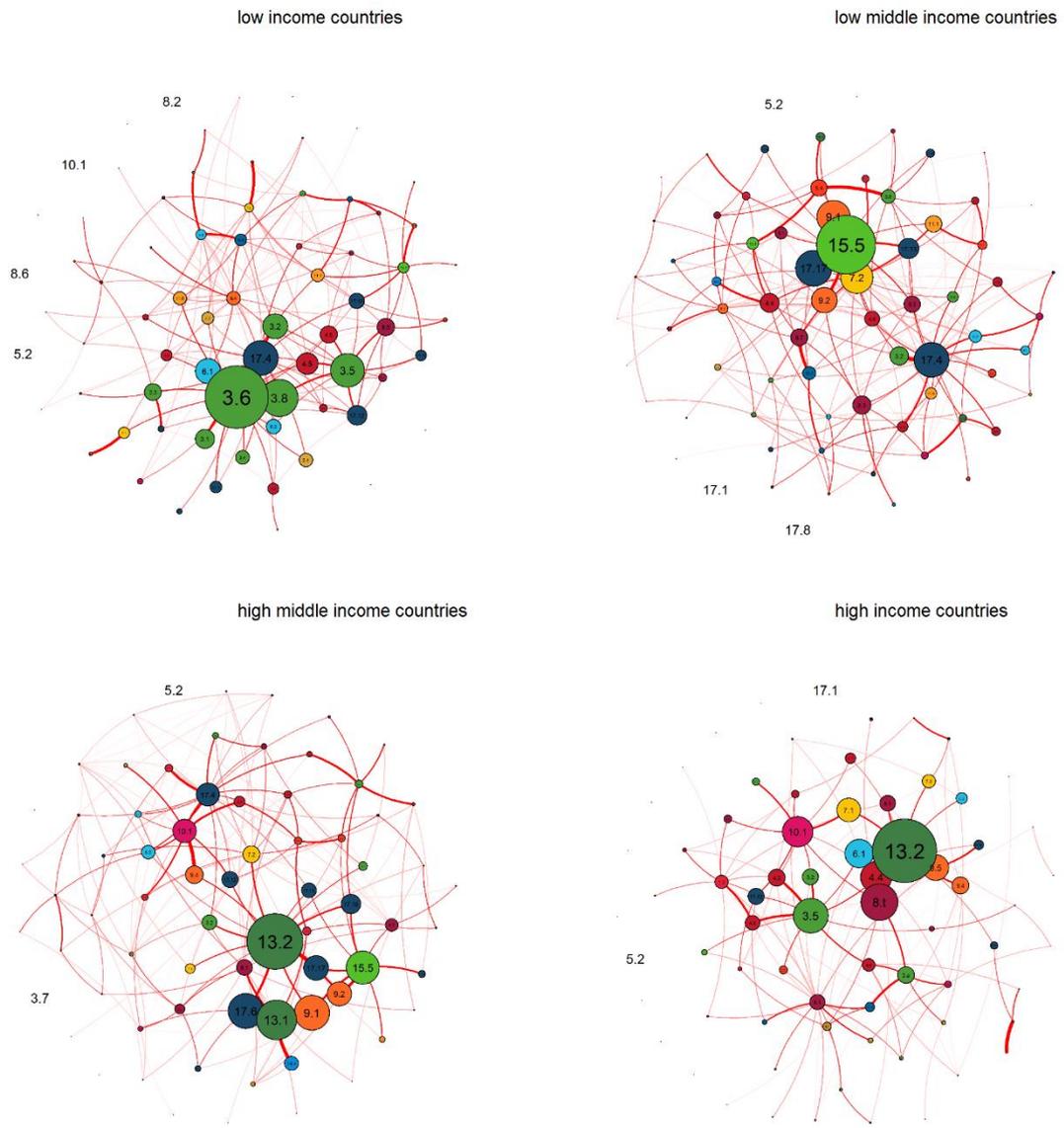

Fig. S7. Subsets of the target sustainomes retaining only negative edges (**A**<0) to help visualise the main barriers in target interactions. Edges range [-1 ; 0[ represented by line thickness. Node size is the eigenvector centrality of targets in this subset sustainome.